\documentclass{article}

\usepackage[nonatbib, preprint]{neurips_2024}

\usepackage[utf8]{inputenc} % allow utf-8 input
\usepackage[T1]{fontenc}    % use 8-bit T1 fonts
\usepackage{hyperref}       % hyperlinks
\usepackage{url}            % simple URL typesetting
\usepackage{booktabs}       % professional-quality tables
\usepackage{amsfonts}       % blackboard math symbols
\usepackage{nicefrac}       % compact symbols for 1/2, etc.
\usepackage{microtype}      % microtypography
\usepackage{xcolor}         % colors
\usepackage{lipsum}

\usepackage{amssymb}
\usepackage{amsmath}

\usepackage{bm}

\usepackage{graphicx}
\graphicspath{ {./images/} }

\usepackage{caption}
\usepackage[numbers]{natbib}
\usepackage{ulem}

\usepackage[disable]{todonotes}

\title{Manifold Transform by Recurrent Cortical Circuit Enhances Robust Encoding of Familiar Stimuli
}

\author{%
Weifan Wang$^{1, 3}$ \quad Xueyan Niu$^2$ \quad Tai-Sing Lee$^3$ \\
$^1$Department of Biomedical Engineering, Carnegie Mellon University \\
$^2$Center for Neural Science, New York University \\
$^3$Center for Neural Basis of Cognition, Carnegie Mellon University \\
\texttt{tai@cnbc.cmu.edu}\\
}

\begin{document}

\maketitle

\begin{abstract}
    A ubiquitous phenomenon observed throughout the primate hierarchical visual system is the sparsification of the neural representation of visual stimuli as a result of familiarization by repeated exposure,  manifested as the sharpening of the population tuning curves and suppression of neural responses at the population level. In this work, we investigated the computational implications and circuit mechanisms underlying these neurophysiological observations in an early visual cortical circuit model. We found that such a recurrent neural circuit, shaped by BCM Hebbian learning, can also reproduce these phenomena. The resulting circuit became more robust against noises in encoding the familiar stimuli. Analysis of the geometry of the neural response manifold revealed that recurrent computation and familiar learning transform the response manifold and the neural dynamics, resulting in enhanced robustness against noise and better stimulus discrimination. This prediction is supported by preliminary physiological evidence. Familiarity training increases the alignment of the slow modes of network dynamics with the invariant features of the learned images. These findings revealed how these rapid plasticity mechanisms can improve contextual visual processing in even the early visual areas in the hierarchical visual system.
\end{abstract}

 \section{Introduction}
Familiarity suppression refers to a phenomenon observed in the inferotemporal cortex (ITC) \cite{meyer2014image, fahy1993neuronal, xiang1998differential, mruczek2007context, sobotka1993investigation, freedman2006experience, woloszyn2012effects} and more recently in early visual cortex \cite{huang2018neural}  that repeated exposure to a set of familiar visual stimuli will lead to the suppression of neural responses to these stimuli, particularly in the later part of the temporal responses. There is evidence in the inferotemporal cortex that familiarity training leads to the sparsification of population neural representation to the familiar stimuli, as neurons' responses to their preferred familiar stimuli were found to be enhanced, even though their responses to their non-preferred familiar stimuli were suppressed, resulting in a sharpening of the stimulus selectivity tuning curves \cite{woloszyn2012effects, freedman2006experience}. It has been inferred from these findings that the BCM Hebbian learning rule is involved in mediating these effects\cite{lim2015inferring}. 

In the early visual cortex, Huang et al. \cite{huang2018neural} showed that neurons with localized receptive fields became sensitive to the global context of familiar images. Based on timing, it can be inferred that this sensitivity is mediated by the recurrent circuits within V2 rather than feedback from higher visual areas. The findings suggest a rapid plasticity mechanism in the early visual cortex modifying the recurrent circuit to encode global or semi-global familiar image context within each visual area along the visual hierarchical system, and not just in ITC. Nevertheless, the computational mechanisms and rationales of such a rapid plasticity mechanism are not well understood.  In this paper, we develop a V1-based neural circuit model, based on BCM Hebbian learning as well as other standard V1 circuitry, that can account for the familiarity training effects. We simulated and analyzed this circuit to show that familiarity training gives rise to a transformed recurrent circuit that is more robust against noises and has better stimulus discriminability for familiar images. This observation is supported by data from a preliminary neurophysiological experiment. We found that cortical recurrent circuit improves robustness against noise by making the slow modes of the network dynamics more aligned with the invariant features of the encoded stimuli. Our study provides a novel manifold transform perspective on cortical recurrent circuits as well as insights into the functional rationales underlying the familiarity learning observed in the early visual cortex.

\section{Recurrent neural circuit model of visual cortex shaped by BCM learning}
\label{sec:model}
\begin{figure}[h!]
    \centering
    \centerline{\includegraphics[width=\linewidth]{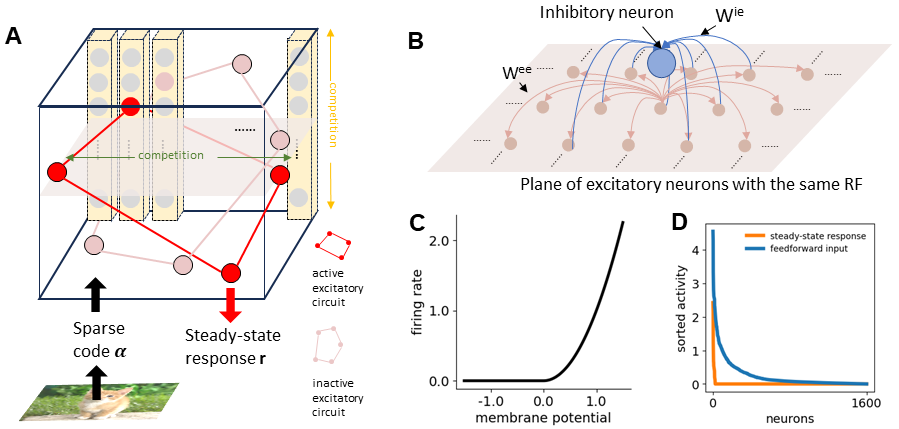}}
    % \captionsetup{width=\textwidth}
    \caption{Recurrent circuit model of the visual cortex. (A) Illustration of the network structure. An excitatory circuit would be formed between excitatory neurons in different hypercolumns, and only certain circuits could be activated given a particular image. Competitions exist between neurons with the same receptive field or in the same hypercolumn. The input to the network is the sparse code $\bm{\alpha}$ of the image and the output is its steady-state population response $\bm{r}$. (B) Illustration of the excitatory neighborhood (for a single RF) and inhibitory neighborhood. (C) Squared-relu activation function. (D) Sorted population activity to an example image in feedforward input and steady-state response.}
    \label{fig:network}
\end{figure}

Familiarity training effects have been reported in macaque ITC and V2. Similar effects have also been observed by us in macaque V1, as well as in mouse V1 (\cite{aitken2024simple, cooke2015visual, kaplan2016contrasting}). Since V1 circuits are relatively well understood and modeled, we decided to use the V1 model for this study, believing that the circuit mechanism is canonical and can be generalized to all visual areas along the ventral visual hierarchy. We constructed a neural circuit model of the primary visual cortex to demonstrate that the plastic horizontal connections, driven by the BCM rule \cite{cooper2012bcm, law1994formation, lim2015inferring}, can reproduce familiarity effects. %Our approach involves constructing a circuit model inspired by V1 architecture, capable of learning horizontal excitatory connections between neurons from the natural scene by leveraging the BCM rule \cite{cooper2012bcm, law1994formation}. 
The network model (fig.\ref{fig:network}A) is a firing-rate-based recurrent neural network with $N_h$ hypercolumns (with $N_r$ rows and $N_c$ columns). Each hypercolumn comprises $N_k$ excitatory neurons with receptive fields (RF) derived from sparse coding \cite{olshausen1996emergence}. We have $N_e= N_r \times N_c\times N_k$ excitatory neurons and the same number ($N_i$) of inhibitory neurons in the network.

Each excitatory neuron $i$ receives projection from its excitatory neighborhood ($NE(i)$) with a range $R_e$ that spans all RFs (fig.\ref{fig:network}B). Each inhibitory neuron $i$ receives projection from the excitatory neurons with the same RF located in its inhibitory neighborhood ($NI(i)$) with range $R_i$ (fig.\ref{fig:network}B) and projects back to all excitatory neurons in the network, forming iso-orientation suppression. We have $|NE(i)|= N_k\times(2R_e+1)^2$, and $|NI(i)|=(2R_i+1)^2$ for neurons not at the border. Excitatory neurons within the same hypercolumn are subject to subtractive normalization governed by an inhibitory neuron that receives the input from all neurons in that hypercolumn and is projected back to all those neurons.

The dynamics of the excitatory population and inhibitory population are written respectively as:
\begin{gather}
    \tau_e \frac{dr^e_i}{dt} = -r^e_i + \sigma(\sum\nolimits_j W_{ij}^{ee}\, r^e_i - \sum\nolimits_{j'} W_{ij'}^{ei}\, r^i_{j'} + \gamma I^e_i) \label{1} \\
    \tau_i \frac{dr^i_i}{dt} = -r^i_i + \sigma(\sum\nolimits_j W_{ij}^{ie}\, r^e_j) \label{2}
\end{gather}
where $r^e_i$, $r^i_i$ is the firing rate of the $i^{th}$ excitatory neuron and inhibitory neuron, respectively. $W_{ij}^{ee}$ is the E-E connection from excitatory neuron $j$ to excitatory neuron $i$; similarly $W_{ij}^{ie}$ and $W_{ij}^{ei}$ are the E-I and I-E connection; $I^e_i$ is the sparse code input to the excitatory neuron $i$. $\gamma$ is the scaling factor controlling the saliency of the input. $\sigma(.) = \lfloor . \rfloor_+^2$ is the squared relu activation function (fig.\ref{fig:network}C). $\tau_e$ and $\tau_i$ are the time constants of excitatory and inhibitory groups.

We assume that only $W^{ee}$ is learnable, while $W^{ei}$ and $W^{ie}$ are fixed. 
%Note that the $W^{ei}$ exists everywhere, but $W^{ee}_{ij} = W^{ee}_{ji} = 0$ for $j \notin NE (i)$ and $W^{ie}_{ij} = W^{ie}_{ji} = 0$ for $j \notin NI(i)$. 
The initial value of the E-E connection ($W^{ee,I}_{ij}$) is constant wherever it exists and equals $w_{ee}/|NE(i)|$. Similarly, the value of E-I and I-E connection are $w_{ie}/|NI(i)|$ and $w_{ei}/N_i$. The BCM update rule \citep{law1994formation} is written as:
\begin{gather}
    \tau_w\, \frac{dW^{ee}_{ij}}{dt} = r_j^e\, r_i^e\, \frac{r_i^e - \theta_i}{\theta_i}; \quad
    \tau_\theta\, \frac{d\theta_i}{dt} = -\theta_i + (r^{e}_i)^2 
\end{gather}
where $\theta_i$ is the BCM threshold for excitatory neuron $i$, $\tau_w$ is the synaptic time scale and $\tau_\theta$ is the homeostatic time scale. Furthermore, we added synaptic scaling to preserve the total strength of pre-synaptic connections to excitatory neuron $i$ throughout the learning:
\begin{equation}
    \sum\nolimits_{j=1}^{N_e}\, W_{ij}^{ee} = |NE(i)|\, W^{ee,I}_{ij}
\end{equation}

\begin{figure}[h!]
    \centering
    \centerline{\includegraphics[width = \linewidth]{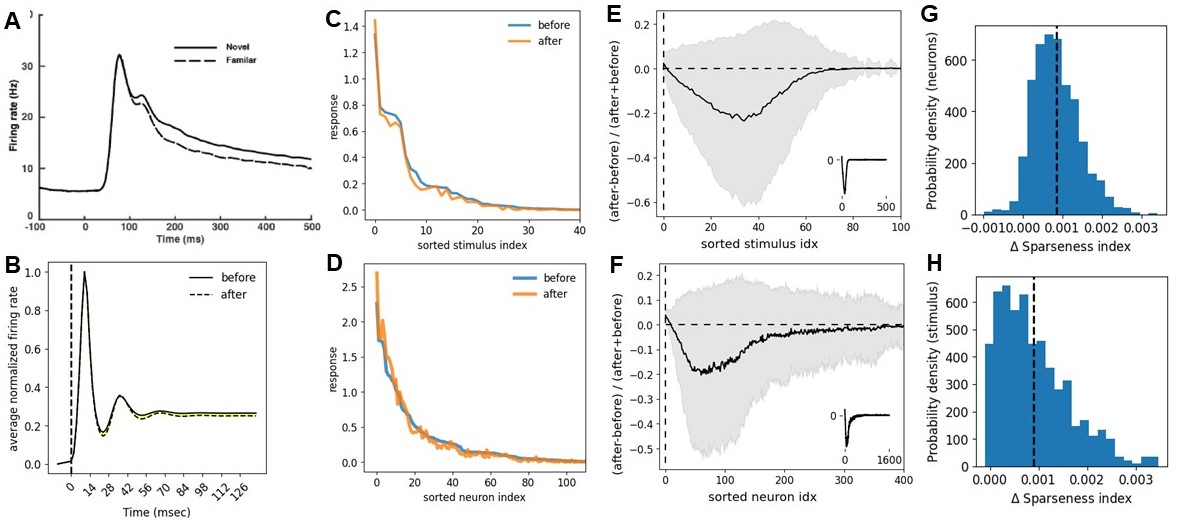}}
    % \captionsetup{width=.85\paperwidth}
    \caption{Tuning curve sharpening and familiarity suppression. (A) Familiarity suppression in primate visual cortex. Figure adapted from \citet{huang2018neural}. (B) Familiarity suppression in the model. The yellow area indicates the reduction after training. (C) Example stimulus tuning curve of one neuron. (D) Example population tuning curve to one stimulus. (E) Effect of familiarity training on stimulus tuning curves quantified by the relative difference ((after-before)/(after+before)). The solid black curve represents the mean, and the grey area represents the std for neurons. The X-axis is the stimulus index of the pre-trained tuning curves sorted in descending order. The main plot only shows the first 100 stimuli, while the inset shows the mean for all stimuli. (F) Similar to E but for the population tuning curve. (G, H) Histogram of the relative difference in the sparseness index of stimulus and population tuning curves.}
    \label{fig:fam_sup}
\end{figure}

%\section{Familiarity suppression and tuning curve sharpening in the model}
%\label{sec:fam}
\paragraph{Validation of the Model} We verified that the model could indeed reproduce familiarity effects, i.e., familiarity suppression and tuning curve sharpening. Fig.\ref{fig:fam_sup}B shows the averaged peri-stimulus firing histogram (PSTH) across images and neurons. Neural activity in the network peaks right after stimulus presentation, an effect of the feedforward input's injection. The initial peak is followed by a bump of activity caused by the surrounding excitation from the neurons in the excitatory neighborhood. Finally, the activity goes into a steady state. The network profoundly sparsifies the population response (fig.\ref{fig:network}D) as a result of the strong surround inhibition and supra-linear activation function. 

We quantified the effect of familiarity training, or change in sparseness or sharpness of the stimulus tuning curves of neurons (fig.\ref{fig:fam_sup}C) using the relative difference index (after - before) / (after + before). The neuron index was obtained by sorting the pre-trained response in descending order. Responses of individual neurons to their highly preferred stimuli were changed by the BCM learning of recurrent connections. Among these stimuli, learning suppressed its response to most of the stimuli while only increasing the response to stimuli with the highest affinity (fig.\ref{fig:fam_sup}E), which indicates that a neuron's tuning curve becomes sharper with enhanced responses to its highly preferred familiar stimuli but with suppressed responses to most of the other familiar stimuli. This is confirmed by the increased sparseness index after training, indicating a more long-tailed tuning distribution (fig.\ref{fig:fam_sup}G).

A similar procedure was used to characterize the change in the population tunings to stimuli (fig.\ref{fig:fam_sup}D). Familiarity training results in activity change in about 10\% of neurons responding to a single image, and most of those neurons are suppressed. In contrast, only a small portion of the neurons responding strongly to the familiar image are enhanced (fig.\ref{fig:fam_sup}F). The population response became more selective and long-tailed as well (fig.\ref{fig:fam_sup}H). Since most neurons are suppressed, the averaged activity of all neurons shows reduction (fig.\ref{fig:fam_sup}B), leading to the familiarity suppression effect. These findings suggest that BCM learning of excitatory recurrent connections is sufficient to produce the familiarity suppression and tuning sharpening found in the primate visual cortex (fig.\ref{fig:fam_sup}A).

\section{Manifold transform in the recurrent cortical circuit}
\label{sec:theory}
In this section and the following, we investigate the computational rationales underlying the learning of the circuit that produces the familiarity effects. We propose to consider the recurrent circuit as performing a manifold transform to improve the representation of the familiar stimuli in terms of efficiency, robustness, and discriminability. Let us consider the problem from a manifold perspective. A given image is represented or encoded by the populational vector of the network's steady-state response $\bm{r}$ to it. Consider a collection of representations of a set of images that are related to each other, and suppose a smooth, continuous transformation underlies the generation of the image set. Such a collection is thought to form a low-dimensional manifold, called neural response manifold \cite{chung2021neural, kriegeskorte2021neural},  in the high-dimensional neural space, where each image will have its unique location. The geometry of the manifold determines what encoded information in the network is available for the downstream decoder. Both biological and artificial neural systems have been shown to transform the response manifold to some specific desired geometry in order to accomplish certain perceptual goals \citep{dicarlo2007untangling, ansuini2019intrinsic}.

\begin{figure}[ht!]
%\vspace*{-0.5in}
    \centering
    \centerline{\includegraphics[width=\linewidth]{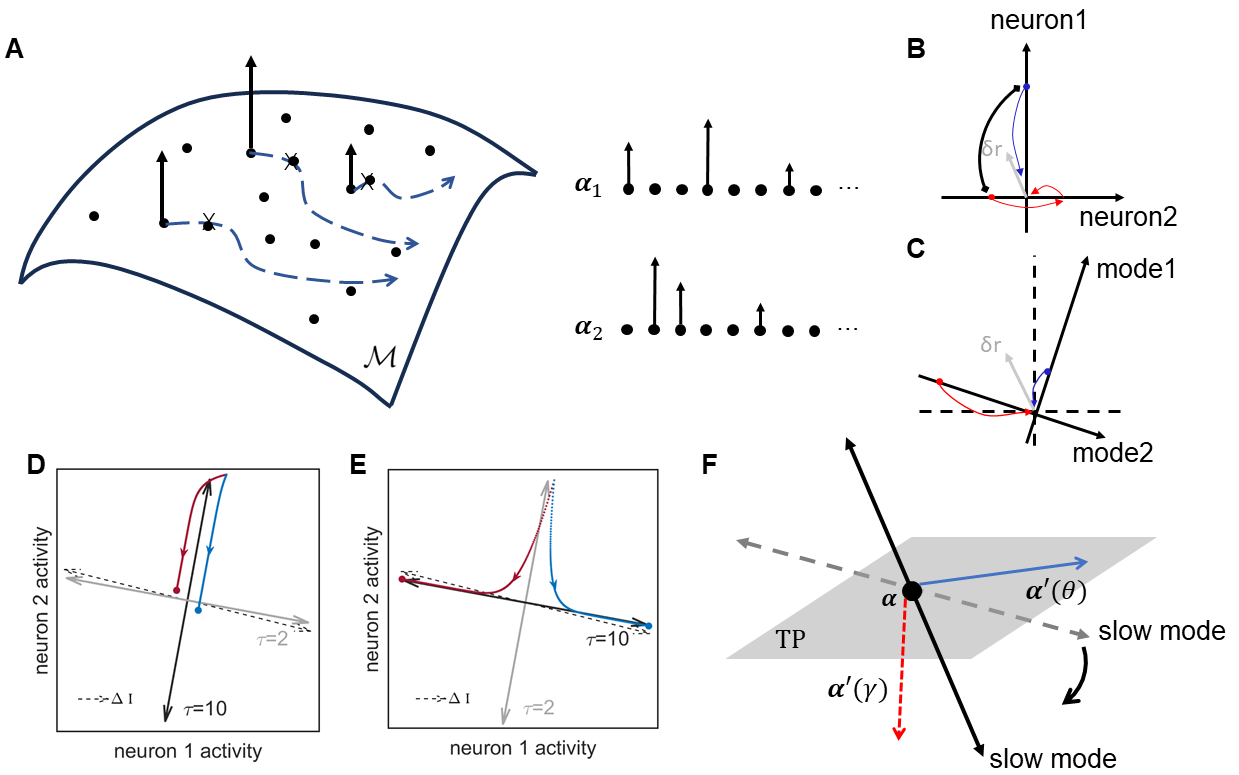}}
    % \captionsetup{width=.85\paperwidth}
    \caption{Manifold transform in the recurrent circuit. (A) Conceptual model illustrating how sparse code abrupts inherited geometric structures in images. $\mathcal{M}$ is the putative image manifold; black dots represent sparse dictionaries that are landmarks on the manifold; black arrows are delta functions on the manifold domains constituting the sparse code of an image; blue arrows show the direction of flows; $\bm{\alpha}_1$, $\bm{\alpha}_2$ are two example sparse codes of two images along the flow. (B) Neuronal response to the small perturbation $\delta r$ around the fixed point (origin). Blue and red arrows illustrate two connected neurons decaying back to the fixed point. (C) Projections of the neuronal activity onto fast-decaying mode and slow-decaying mode. The projection decouples the neuronal dynamics, leading to independent decay towards the fixed point (origin). (D-E) Illustration of different system behaviors when the slow mode is aligned or orthogonal to the input difference (dashed arrow). Blue and red curves are activity traces driven by two constant inputs. Black arrow: slow decaying mode, grey arrow: fast decaying mode. Invariant features are in the orthogonal direction of the input difference. (F) Illustration of how the recurrent circuit performs manifold transform. TP: tangent plane of the sparse code manifold; $\bm{\alpha}'(\theta)$: derivative of sparse code w.r.t. $\theta$; $\bm{\alpha}'(\gamma)$: derivative of sparse code w.r.t. $\gamma$. Training should render slow modes of the network dynamics more orthogonal to $\bm{\alpha}'(\theta)$ than to $\bm{\alpha}'(\gamma)$ (i.e. more aligned with the invariant features w.r.t. $\theta$ than to $\gamma$), leading to neural responses that are more similar to the neighbor stimuli compared to other stimuli.}
    \label{fig:theory}
\end{figure}

Studies on manifold learning \citep{roweis2000nonlinear, wiskott2002slow, chen2018sparse} have long noticed that the primary sensory areas, such as V1, with neurons selective to local features, such as the sparse code dictionary used in our model, cannot represent the geometry inherited in images. Smooth transitions in images are often accompanied by abrupt changes in the population neural codes.
%To see how that happens, consider the following conceptual model with a manifold $\mathcal{M}$ capturing the true geometry relations between stimuli. 
From a manifold perspective, as discussed \citet{chen2018sparse}, the learned spare code dictionaries are discrete samples on the manifold $\mathcal{M}$, and the sparse code $\bm{\alpha}_i$ of a stimulus $\bm{x}_i$ is a composition of multiple delta functions on the manifold domain (fig.\ref{fig:theory}A). The underlying transformation of the images, such as those induced by changing noise level, changing view angle, etc., is a collection of flows on $\mathcal{M}$. Because the learned dictionaries are typically unordered, the flow on the manifold would lead to a drastic change in the sparse code (fig.\ref{fig:theory}A). An objective of manifold transform is to search for the manifold $\mathcal{M}$ that captures the true geometric relationship between stimuli given the sparse code. % {\bf Should we cite Chen and Olshausen here}

The traditional manifold learning approaches can be leveraged to solve this manifold searching problem \citep{roweis2000nonlinear, chen2018sparse}. Specifically, a transform $g(\cdot)$ is learned to preserve the local geometry determined by the distance or similarity between stimulus $\bm{x}_i$ and its local neighbors $\bm{x}_{n(i)}$. Here, we define similarity as the first-order derivatives, assuming the actual change induced by the transformation is small compared to the change in the sparse code so that similar stimuli will be close together in this manifold. %clustered together stimulus within the neighborhood should be clustered together. 
This can be obtained by minimizing, as objective, the first-order derivatives within the neighborhood while maintaining the variance of stimulus representations to prevent degenerative solutions. 
%\citet{chen2018sparse, chen2022minimalistic} proposed the sparse manifold transformation (SMT) where the transform are multiple linear pooling functions learned by maximizing similarity while constraining representation variance to be uniform \citep{chen2018sparse}. 
For the connectivity ${\bm W}$ of the circuit, this objective can be relaxed to the ratio between the magnitude of two derivatives:
\begin{equation}
    \min_{{\bm W}} \sum_i\, \frac{\| \bm{\beta}'(\theta_i) \|^2}{\| \bm{\beta}'(\gamma_i) \|^2},\quad \text{s.t.}\quad \bm{\beta} = g(\bm{\alpha}; {\bm W}),
\end{equation}
where $\theta$ controls the smooth variation to local neighbors, such as noises and view angles, and $\gamma$ controls the variation to other non-local neighbors such as images of different objects. This objective keeps within-neighborhood variation small relative to other variations outside of the neighborhood.

What are the desired neural dynamics of the recurrent circuit that implements the manifold searching? The objective (Eq.6) implies that the desired transform $g(\cdot)$ manipulates local variations around each point on the manifold to effectively alter the global geometry of the manifold, an approach termed 'think globally, fit locally' \citep{saul2003think, chen2022minimalistic}. Therefore, it is appropriate to analyze the local dynamics of the network around each fixed point to understand how the recurrent circuit implements manifold searching. The dynamics of small perturbations around a particular fixed point can be approximated by a linear dynamical system, and the linearized dynamics can be decomposed into distinct activity patterns (modes), each unfolds with a different dynamical timescale $\tau$ and filter or project the input signal in a different way \cite{sussillo2013opening, chadwick2023learning, goldman2009memory, mante2013context, maheswaranathan2019reverse}. Specifically, each dynamical mode describes a specific deviation pattern of population activity from the fixed point, which evolves independently (fig.\ref{fig:theory}B-C) with speed controlled by the respective decaying time constant $\tau$. A negative $\tau$ entails escaping from the fixed point, while a positive $\tau$ entails decaying back to the fixed point. Mathematically, projecting the perturbed population activity onto a specific mode results in the independent, exponential growth or decay of the projected activity (see Appendix B).
	
Similar to the local dynamics, the tuning derivative of the response vector relative to the stimulus parameter, $\bm{r}' = \partial \bm{r} / \partial \theta$, evaluated at certain $\theta$ can also be decoupled into disentangled components that fall along each dynamical mode. Specifically, the derivatives at stimulus parameter $\theta$ of the projected response onto a set of dynamical modes $L$ are independently and linearly scaled with the projection of the input derivatives $\bm{\alpha}'$  at $\theta$ onto corresponding modes:
\begin{equation}
    L^T\, \bm{r}'(\theta_i) = {\bm T} \Tilde{L}^T \bm{\alpha}'(\theta_i),
\end{equation}
where diagonal matrix ${\bm T} = \delta_{ij} \tau_i$ containing decaying time constants of each mode, $\bm{\alpha}$ is the sparse code input to the recurrent circuit (see Appendix B), the projected steady-state response of the recurrent network $L^T \bm{r}$ is $\bm{\beta}$ (from Eq 6), the output of the manifold transform $g(\cdot)$. Eq.7 implies if slowly-decaying dynamical modes of the network extract the features in the input that are invariant w.r.t. $\theta$ (i.e. orthogonal to $\bm{\alpha}'(\theta_i)$), then the $\bm{\beta}$ representation of the stimuli associated with the same invariant features will become closer together (fig.\ref{fig:theory}D-E).

Minimizing the derivative ratio (eq.6) will cause input features aligned with the slow modes to change more slowly w.r.t. $\theta$ than with $\gamma$ around each fixed point. This is achieved by making the slow modes of the network dynamics aligned more to the invariant features w.r.t. $\theta$ over those w.r.t. $\gamma$ (fig.\ref{fig:theory}F). Then the global geometry of the manifold can be recovered by locally optimizing the similarity of $\bm{\beta}$ representations of the stimuli within the neighborhood.

The projections of the network activity $r$ on the slow modes provide an optimal linear readout that reflects the manifold geometry. Projection of network activities onto an arbitrary vector will contain a superposition of different input projections (eq.12 in Appendix B), which is suboptimal for extracting invariant features. The output of a recurrent network allows the projection to the slow modes to be selectively read out by the downstream neurons, which are optimal for manifold geometry. The projection to the slow modes allows the specific invariant features to be optimally read out by downstream neurons from the network responses. 

%Since some projections of the input identify the invariant features while others do not, the arbitrary projection of recurrent network activity is the superposition of different input projections (eq.12 in Appendix B), which is sub-optimal.

%The dynamical modes, on the other hand, extract single input projections and retain the feature in the corresponding response projection.

\section{Transformation of neural response manifold to improve noise robustness}
\label{sec:manifold}

\begin{figure}[h]
\vspace*{-0.00in}
    \centering
    \centerline{\includegraphics[width = \textwidth]{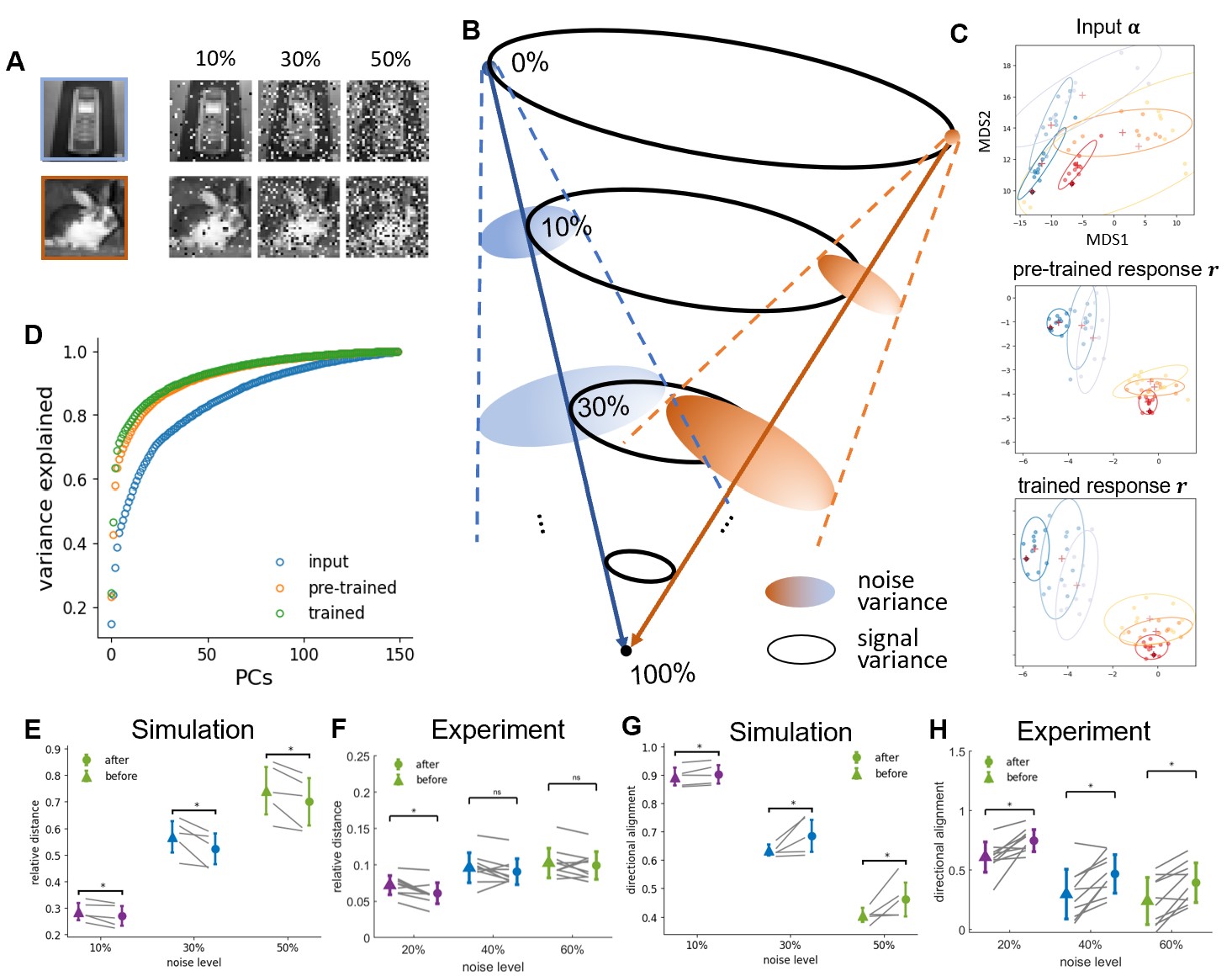}}
    % \captionsetup{width=.85\paperwidth}
    \caption{Manifold transformation in the noise experiment. (A) Example images corrupted with different levels of noise. (B) Cartoon illustration of the cone formed by decreasing signal variance and two cones formed by increasing noise variance of two example images. (C) Visualization of the manifold in 2D formed by fixed points corresponding to all noise-corrupted samples of 2 example images. Each figure represents the manifold at different stages of manifold transformation. Input: feedforward input; pre-trained: pre-trained network; trained: epoch5 trained network. Two color gradients delineate two example images, and deeper colors represent lower noise levels. Dots: fixed points; Crosses: cluster centroids; Ellipses: cluster covariance. Pre-trained and trained figures have the same scale on the x and y-axis. (D) Fraction of variance explained of different PCs for feedforward input and fixed points of pre-trained and epoch5 trained networks. (E-F) Relative distance of noise images with clear targets. (E) was obtained from model simulation, and (F) was obtained from primate neural data. Triangle marker: before training, circle marker: after training (epoch5 for the simulation). Grey straight lines are for individual images. *: p<0.05. Error bar represents std across images. (G-H) Similar to a-b but for the directional alignment.}
    \label{fig:manifold}
\end{figure}

We have verified in Section 2 that familiarity training leads to greater sparsity in stimulus tuning and population code, hence making the neural codes for familiar stimuli more efficient. Here, we performed a simulation experiment to study whether the trained circuit becomes more robust against noise as a consequence of the transformation of the neural response manifold. During familiarity training, a set of 10 images, each corrupted by random salt and pepper noises of 0\%, 10\%, 30\%, and 50\%, are repeatedly presented to the network. In each epoch, each stimulus and its noisy versions are presented and updated at least 150 times according to the BCM learning. Fig.\ref{fig:manifold}A shows examples of two images and their versions corrupted by different degrees of noise. Conceptually, different stimuli lie on a ring in high-dimensional space. Corrupted by noises, the different stimulus groups (clear and noisy versions) for distinct rays converge to the 100 \% noise center in a cone shape (fig.\ref{fig:manifold}B) 
%We study the neural response manifold using the setup of the noise experiments that examine whether familiarity training would increase the robustness of neural representation against noise in the image. During training, a pair of images A and A' were given to the network, where A' is the noise-corrupted version of A. A clear image would be corrupted by different noise levels (10\%, 30\%, and 50\%), and each noise level is associated with 10 noise patterns (Fig.\ref{fig:manifold}A). The network is trained for 5 epochs, where all clear-noise pairs were presented to the network once. The desired representation should be robust against noise, such that the encoding of noise-corrupted image A' should be more similar, or geometrically closer to the corresponding clear image A (target) than to other clear images.

After each epoch, we probed the neural activity and analyzed the neural response manifold formed by fixed points of the excitatory population. We found that the configuration of the neural response manifold is consistent with the conceptual model (fig.\ref{fig:manifold}B, also see the visualization in Appendix C). We investigated the transformation of the manifold from input $\bm{\alpha}$ to steady-state response $\bm{r^e}$, and from pre-trained response to trained response. Firstly, The input (sparse code) has much higher PC dimensions than the steady-state response (fig.\ref{fig:manifold}D), which suggests that recurrent computation constrains the manifold in a lower dimensional space. Second, the familiarity training also slightly reduces the manifold's dimension, but the magnitude is low compared to that caused by recurrent computations. Fig.\ref{fig:manifold}C visualizes the manifold transformation for 2 example images by embedding manifolds at different stages of computation or different training epochs into the same subspace. The transformation caused by recurrent computation largely compresses the manifold and disentangles the manifold of two images. In contrast, the transformation induced by familiarity training expands two manifolds with a low magnitude and involves more nuanced changes. 

We then asked whether the transformation facilitates noise robustness by quantitatively measuring the directional alignment and the relative distance to the target. The relative distance measures how close a noisy image is to the target than to other incorrect clear images. We found that for all noise levels, the directional alignment increases and the relative distance decreases, either by comparing before/after recurrent computation or comparing different epochs (fig.\ref{fig:manifold}D, F). 
We carried out a neurophysiological experiment recording from 70 V1 and V2 neurons of a macaque monkey using a Gray-Mater SC96 array implanted over the V1 opeculum. 12 V1 and 38 V2 neurons can be tracked using the fingerprint stimuli across three days. Each day, each of the 10 images is presented at least 400 times in the clear version and 100 times in each of the three levels of salt-pepper noise corruption. (All experimental protocols have been approved by the Institutional Animal Care and Use Committee of our university). Fig.\ref{fig:manifold}E, G compared the directional alignment and relative distance to the steady-state responses of the clear images on the second day and the fourth day of training. Trends similar to the model's predictions (fig.\ref{fig:manifold}D, F) can be observed: noisy images at all noise levels become more similar to the steady state responses of their 0\% noise target image relative to other clear images, showing that the rapid plasticity in the recurrent networks indeed enhanced the robustness of neural representation against noise, as a consequence of the modification of the neural response manifold due to familiarity training. \todo{redo fig 4 layout, correct legend}

\section{Familiarity learning aligns invariant stimulus features with the slow modes of network dynamics}
\label{sec:dynamic}
\begin{figure}[h!]
    \centering
    \centerline{\includegraphics[width=\textwidth]{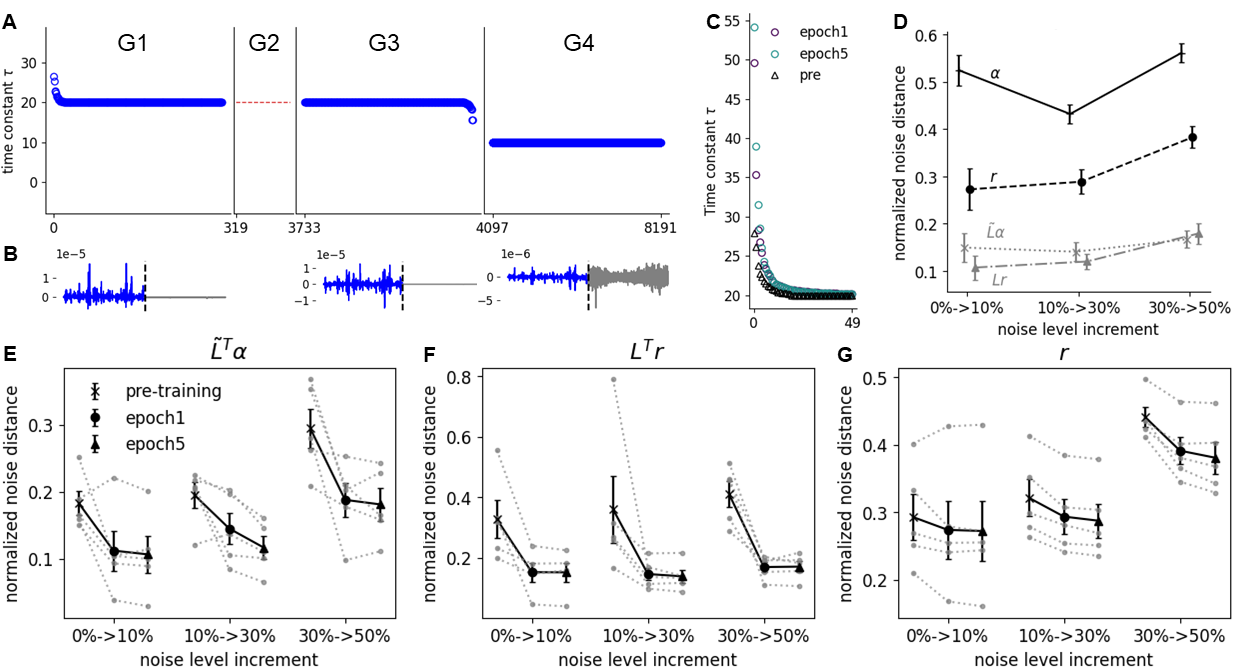}}
    % \captionsetup{width=.85\paperwidth}
    \caption{Familiarity training aligns slow dynamical modes with invariant features. (A) Groups of dynamical modes that are specified by decaying the time constant $\tau$ obtained by linearizing around a specific fixed point. First panel: Group 1, with $\tau > \tau_e$; Second panel: Group2, with $\tau = \tau_e$ and all eigenvectors are zeros vectors. Third panel: Group 3, with $\tau_i < \tau < \tau_e$. Last panel: Group 4 with $\tau = \tau_i$. (B) Averaged eigenvectors of groups 1, 2, and 4 correspond to excitatory neurons (colored in blue) and inhibitory neurons (colored in gray). (C) Time constants of the top 50 slow decaying modes, averaged across images and noise levels, for pre-training, epoch1-trained, and epoch5-trained networks. (D) Normalized noise distance in the epoch5 trained network. $\alpha$: feedforward input; $r$: neural response; $\Tilde{L}^L r$: input projection; $L^T r$: response projection. Error bar represents s.e.m. across images. (E-G) Training effects of normalized noise distances regarding input projection, response projection, and original neural response. Error bar represents s.e.m. across images. Gray lines delineate single images.}
    \label{fig:dynamic}
\end{figure}

To understand how the circuit carries out the manifold transform,  we investigate whether the recurrent circuit with familiarity training transforms the manifold using the mechanism described in section \ref{sec:theory}. We used numerical methods to perform linearization analysis around each fixed point of the network. The dynamical modes around each fixed point are all decaying, suggesting that all the fixed points are stable attractors. The decaying modes can be categorized into several groups according to the decaying time constants. The first group of decaying modes has time constants greater and gradually approaches $\tau_e$; The second group has $\tau = \tau_e$, but the modes are all zero vector, thus will not contribute to the dynamic; The third group has time constants smaller than $\tau_e$ and larger than $\tau_i$, and the last two groups have $\tau = \tau_i$, decaying with the same speed as the inhibitory neurons (fig.\ref{fig:dynamic}A). Eigenvectors in groups 1 and 3 constitute exclusive loading for excitatory neurons. Hence, they are relevant to the manifold formed by population responses of excitatory neurons. In contrast, eigenvectors in group 4 constitute mainly inhibitory loading (fig.\ref{fig:dynamic}B). These observations are consistent with the finding that the manifold of excitatory neurons is low-dimensional (fig.\ref{fig:manifold}H) since the majority of modes (group 2) do not contribute to the dynamic. Familiarity training increases the decaying time constant of slowest modes for all noise levels of all images, and the increasing effect reduces as the mode's time constant approaches $\tau_e$ (fig.\ref{fig:dynamic}C).

In the noise experiment setup, the local neighbors of a particular stimulus are those belonging to the same image but corrupted by different levels of noise, while stimuli belonging to other images are not considered local neighbors. Therefore, we examined the alignment of the top 20 slow modes with invariant features w.r.t. both noise levels and images. We used the distance between two adjacent noise levels (noise distance) to approximate the tuning derivative w.r.t. noise. The tuning derivative w.r.t. images evaluated at image A was quantified by averaging the distance from image A to all other images at the same noise level (image distance). We found that the dynamic modes operate as a detector of the invariant feature specific to the noise. This is evidenced by the decrease in the normalized noise distances in both the projected input and the projected response compared to the inputs (fig.\ref{fig:dynamic}D). Looking into the noise distance and image distance individually, we found that for input and response projections, the noise distance and image distance decrease, but the decrease in the former is more pronounced (fig.\ref{fig:dynamic_sup}C, D). We also observed the decrease in the normalized noise distance in the original neural responses (fig.\ref{fig:dynamic}D), which is consistent with the finding of representational distances in fig.\ref{fig:manifold}F. Projecting response onto the slow modes further decreases the normalized noise distance (fig.\ref{fig:dynamic}D), confirming that slow modes constitute an optimal subspace for the noise invariance, better reflecting the underlying geometry of stimuli. We also extend the analysis to other subspaces spanned by different numbers of slow modes in the first group, and these observations are consistent (fig.\ref{fig:dynamic_sup}A, B, see Appendix D).

We then looked at the effect of training by comparing the trained network with the pre-trained network. For subspace spanned by the top 20 slow modes, training decreases the normalized noise distance in input projection, response projection, and original network response around each fixed point (fig.\ref{fig:dynamic}E-G). The results suggest that familiarity training facilitates manifold transformation by continuously improving the local alignment of slow dynamical modes with invariant features w.r.t. noise level around fixed points, making the slow subspace better for the noise invariance. For individual distances, training makes noise levels farther away from each other, but the clear images become far away even faster, effectively reducing the relative distance between noise levels (fig.\ref{fig:dynamic_sup2}C, D). Furthermore, the effects of familiarity training are also largely consistent between different subspace dimensions (fig.\ref{fig:dynamic_sup2}A, B, see Appendix D). Overall, these results demonstrate that recurrent networks detect invariant features that are robust against changes in noise levels, by aligning them with slowly decaying modes. Familiarity training further enhances the detection of invariance features. These properties of the network dynamics facilitate the formation of a neural manifold geometry that improves the representation of familiar stimuli and their robustness against noise. 

\section{Discussion}
\label{sec:discussion}
In this work, we investigated the circuit mechanisms underlying the effects of familiarity training in the early visual cortex. We applied the BCM Hebbian learning rule \cite{cooke2015visual, lim2015inferring} to early visual cortical circuit models, with standard cortical circuit motifs of sparse codes, horizontal excitatory connections, and competitive normalization mediated by inhibitory interneurons to account for the neurophysiological observations of familiarity suppression \cite{huang2018neural} as well as the sharpening of tuning curves in the early visual cortex. This finding shows that local recurrent connections within each visual area are capable of forming Hebbian circuits to encode the relatively global stimulus context of the familiar images. These relatively non-local circuits can stack up together to form a hierarchy of recurrent circuits, processing familiar images in each area and progressively across areas down the ventral visual hierarchy. We showed that such a circuit could perform image denoising, making the neural circuits more robust against noises based on simulations (fig.\ref{fig:manifold}E, G), which is consistent with the results from a preliminary neurophysiological experiment (fig.\ref{fig:manifold}F, H). 

We advanced the novel proposal that the cortical recurrent circuit performs a manifold transform, and familiarity training induces modifications in the stimulus-encoding manifolds. 
%The different stimuli, as well as their noisy versions in different levels of noise, form a cone manifold in a high-dimensional space (fig.\ref{fig:manifold}B). Familiarity training increases the alignment of the neural representation of the noisy version of an image with the clear version of the image, resulting in better discrimination of corrupted images with familiarity training. 
Applying linear analysis around the attractors in this nonlinear network, we showed that the tuning derivatives can be untangled and decomposed into different dynamic modes operating on different timescales. Recurrent dynamics implement a manifold transform that aligns the invariant stimulus features with the slow modes of the network dynamics. Features aligned with the slow modes would change more slowly within the local neighborhood on the manifold, resulting in an increase in similarity among local neighbors, for example, the different representations of the same image corrupted by noises (fig.\ref{fig:theory}F). Empirically, we found that familiarity training decreases the distances in the noise direction relative to the distances in the stimulus direction locally around each fixed point, resulting in more robust representations globally (fig.\ref{fig:dynamic}D-G). 
% Decomposition of recurrent dynamic has been applied to the study of sensory discrimination, decision making, and continuous attractor networks \cite{chadwick2023learning, mante2013context, fung2010moving}, with hand-crafted or task-trained connections. Here, we focus on the learning aspect of the dynamic of the Hebbian circuits, showing that stimulus-filtering by recurrent dynamic due to familiarity training is important for shaping the geometry of the response manifold. 

% limitation: 
Our model and analysis have several limitations. Firstly, the architecture of our model does not follow the detailed biological neural circuit structure with different types of neurons and connectivity. Rather, we incorporate the standard circuit mechanisms of macaque's visual cortex, such as surround excitation/inhibition, normalization, etc., to construct a generalizable circuit model. Second, due to the discrete nature of the stimuli tested, we used distances to approximate derivatives in the noise experiment. This can introduce small high-order term errors in eq.7, but should have minimum impact on the observations and conclusions regarding the distance of both input and output projections. Lastly, there are no obvious cutoff values for the number of modes used in the analysis, as the time constants do not vary significantly within each of the three groups of modes. We have verified that the observed results are consistent across different numbers of group 1 modes, which decay slower than the excitatory neurons and would be most relevant to the manifold formed by steady-state responses of the excitatory neurons.

\bibliographystyle{unsrtnat}
\bibliography{references}  %%% Uncomment this line and comment out the ``thebibliography'' section below to use the external .bib file (using bibtex).

\newpage

\appendix

\section{Experiment details of familiarity suppression and tuning curve sharpening}
In the simulation, we set $N_r = N_c = 5$. The feedforward response is computed using $N_k = 64$ convolutional filter is obtained using the convolutional sparse coding algorithm that jointly minimizes the reconstruction loss and $L1$ norm of the per-hypercolumn activation \cite{szlam2010convolutional}. The strategy improved the coding efficiency and was shown to enhance the characterization of the receptive field of neurons in the primary visual cortex \cite{wu2021complexity}. The radius for mutual facilitation among excitatory neurons is Re = 2, and the radius for iso-orientation suppression is Re = 1. The decaying time constant for the excitatory population is $\tau_e = 20$, and for the inhibitory population is $\tau_i = 10$. Initial weights of excitatory-excitatory connections, excitatory-to-inhibitory connection, and inhibitory-to-excitatory connections are $w_{ee} = 5, w_{ei} = 1, w_{ie} = 30$. An upbound of $W_{ee}$ is set to be 25 to prevent overflow, and the learning could be stabilized through the sliding threshold as well. Finally, BCM-related time constants are $\tau_w = 2e9, \tau_\theta = 1e7$. 

We trained the network on 500 natural images from CIFAR100. Each image is shown to the network for 150 ms in one epoch of training (forward-simulating Euler method for 150 time steps), and the network was trained for 5 epochs. The network was reset to the initial state ($\bm{r}_e = \bm{0}; \bm{r}_i = \bm{0}$) between each image. In training, we set the saliency strength $\gamma = 30$. The initial BCM threshold for each neuron is determined by averaging its responses across all time steps and images in the pre-trained network.

The sparsity index is introduced in \cite{vinje2000sparse}, which is expressed as: \(S = (1 - [(\sum r_i/n)^2) / \sum(r_i^2/n)]) / (1 - (1/n))\), where $r_i$ indicate the population response to $i^{th}$ image. 

\section{Linearization analysis}
Linearization allows the complicated dynamic of the nonlinear system to be understood using an interpretable linear system as long as the linear term dominates the dynamic in the vicinity of the fixed point. \cite{sussillo2013opening}. To perform the linearization, we first express the network dynamic (eq. 1\&2) using the matrix form: 
\begin{equation}
    \bm{T}\frac{\partial \bm{r}}{\partial t} = -\bm{r} + \sigma(\bm{W} \bm{r} + \bm{I}(\theta)),
\end{equation}
where $\bm{r} = [\bm{r_e};\, \bm{r_i}]$, $T_{ij} = \delta_{ij} \tau_i$, and $W = [W_{ee}, W_{ei}; W_{ie}, \bm{0}]$ (We use "," to separate column, ";" to separate row). $\bm{I}(\theta)$ denotes the feedforward input corresponding to a particular stimulus variable $\theta$. The equation has a general form of $\Dot{\bm{r}} = F(\bm{r})$, and the Taylor expansion around fixed point $\bm{r}_{SS}(\theta)$ is:
\begin{align}
    \Dot{\bm{r}_{SS} + \delta \bm{r}} = F(\bm{r}_{SS}) + F'(\bm{r}_{SS}) \delta \bm{r} + \frac{1}{2} \delta \bm{r}^T F''(\bm{r}_{SS}) \delta \bm{r} + \cdots \notag
\end{align}
Since $\bm{r}_{SS}$ is the fixed points, $F(\bm{r}_{SS}) = 0$, and the second order term vanished assume small $\delta \bm{r}$, we can obtain the linearized dynamic of the network around $\bm{r}_{SS}$ by substituting back eq.8:
\begin{equation}
    \Dot{\delta \bm{r}} \approx (\phi' \bm{W} - \bm{T}^{-1}) \delta \bm{r},
\end{equation}
where $\phi'_{ij} = \delta_{ij} \tau_i^{-1} \sigma'(\bm{W} \bm{r}_{SS} + \bm{I}(\theta))$. This matrix quantifies the sensitivity of each neuron's firing rate to small changes in its membrane potential. 

Eq. 9 defines a simple linear system that approximates the dynamic of the original system near a particular fixed point. We define Jacobian $J(\theta) = \phi' \bm{W} - \bm{T}^{-1}$. Note that the dependency of the Jacobian on the stimulus variable will be omitted in the notation for simplicity. Dynamical modes of the linear system can be found by diagonalizing the Jacobian using eigendecomposition. After diagonalizing $J$, the dynamic can be written as:
\begin{equation}
    \Dot{\delta \bm{r}} = \sum_i^N \lambda_i \bm{v}^R_i (\bm{v}^L_i)^T \delta \bm{r},
\end{equation}
where $\bm{v}^R_i$, $\bm{v}^L_i$ denotes $i^{th}$ right/left eigenvector, and $\lambda_i$ refers to the corresponding eigenvalue. Due to the orthogonality between left and right eigenvectors, by projecting $\delta \bm{r}$ onto the left eigenvector, we efficiently 'diagonalize' the dynamics, such that it is decoupled into multiple independent processes:
\begin{equation}
    (\bm{v}^L_i)^T \Dot{\delta \bm{r}} = \lambda_i (\bm{v}^L_i)^T \delta \bm{r}
\end{equation}
In the main text, the dynamical mode refers to an arbitrary left eigenvector $\bm{v}^L$ here. 

The fact that the left eigenvector has the ability to identify the independent components in the dynamic encourages us to study how these components can be linked to the computations performed in the network. specifically the derivatives along the manifold. Estimating derivatives could be done by differentiating the steady-state response $\bm{r}_{SS} = \sigma(\bm{W} \bm{r}_{SS} + \bm{I}(\theta))$ with respect to stimulus variable $\theta$. This gives:
\begin{equation}
    \frac{\partial \bm{r}_{SS}}{\partial \theta} \Bigr|_{\theta=\theta^*}= -\sum_i^N \frac{1}{\lambda_i} \bm{v}^R_i (\bm{v}^L_i)^T \phi' \bm{I}'(\theta^*)
\end{equation}
This equation demonstrates that the derivative along the response manifold is the superposition of input derivatives processed by different modes, scaled by the corresponding time scale $\tau_i = -1 / \lambda_i$. Therefore, projecting the response derivatives onto the left eigenvector reveals how the input derivative falls along each mode, which obtains:
\begin{equation}
    (\bm{v}^L_i)^T\, \frac{\partial \bm{r}_{SS}}{\partial \theta} \Bigr|_{\theta=\theta^*} = -\frac{1}{\lambda_i} (\bm{v}^L_i)^T \phi' \bm{I}'(\theta^*).
\end{equation}

\section{Manifold transformation in noise experiments}
\subsection{Experiment details}
We randomly picked five 32x32 images from CIFAR-100 that have pixel value standard deviation larger than 0.2 as the clear image used in the noise experiment. The images were converted to the grey scale before adding the noise. To corrupt the image by $N\%$, we randomly chose $N\%$ pixels and replaced the pixel value with $\varepsilon \sim \mathcal{U}(0, 1)$. One noise pattern is defined by one set of pixel indexes and one set of random pixel values. We used three noise levels (10\%, 30\%, 50\%) and 10 noise patterns; hence, there are $5 \times 3 \times 10$ clear-noise pairs.

The simulation setup is the same as in A. \todo{training timeline, 5 epoch}

The fixed point was found by forward simulating the Euler method until convergence ($\|\dot{r}\| < 1e^{-8}$) in the probe test period. The trajectory was visualized by reducing the dimension to 2 via PCA. The fixed point manifold was visualized by embedding the manifold in 2D via Multidimensional Scaling (MDS) that preserves the pairwise Euclidean distances between all fixed points of interest. To measure the directional alignment, we computed the cosine similarity given by 
\begin{equation}
    s(\bm{x}^n_{i}, \bm{x}^n_{0}) = \frac{\bm{\hat{r}^e}(\bm{x}^n_{i})^T \bm{\hat{r}^e}(\bm{x}^n_{0})}{|\bm{\hat{r}^e}(\bm{x}^n_{i})| |\bm{\hat{r}^e}(\bm{x}^n_{0})|},
\end{equation}
where $\bm{r^e}(\bm{x}^n_{i})$ refers to steady-state response to image n with noise level i averaged across noise patterns, and $\bm{\hat{r}^e}(\cdot) = \bm{r^e}(\cdot) - \langle \bm{r^e}(\bm{x}^n_{0}) \rangle$ represents the population vector of excitatory neurons that starts from the mean of clear images. The relative distance is computed as: 
\begin{equation}
    d(\bm{x}^n_{i}, \bm{x}^n_{0}) = \frac{\sqrt{(\bm{r^e}(\bm{x}^n_{i}) - \bm{r^e}(\bm{x}^n_{0}))^2}}{\sqrt{\sum_{i \neq n} (\bm{r^e}(\bm{x}^n_{i}) - \bm{r^e}(\bm{x}^i_{0}))^2\, /\, (N-1)}}
\end{equation}

The method for recording and pre-processing of Macaque neural data is similar to \citet{huang2018neural}.

\subsection{Population activity and tunings}
\begin{figure}[ht!]
    \centering
    \centerline{\includegraphics[width=\linewidth]{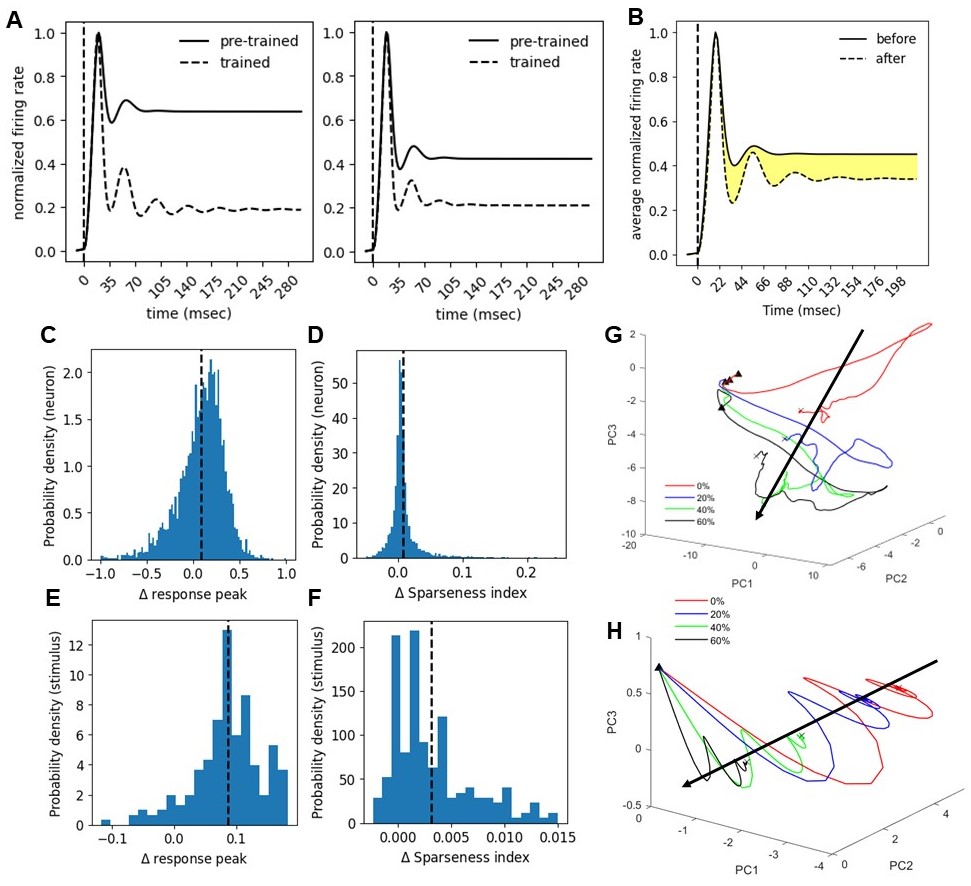}}
    % \captionsetup{width=.85\paperwidth}
    \caption{Population firing rate and neuron tuning in noise experiment. (A) Firing rate histogram of two example neurons averaged over all stimuli. The left neurons have multiple bumps after training, and the right neurons do not show multiple bumps. (B) Averaged firing rate over all stimuli and all neurons for pre-trained and epoch5-trained networks. The yellow area represents the reduction in average firing rate caused by familiarity training. (C-D) Histogram of relative difference ((after-before) / (after+before)) in the response peak and sparseness index for stimulus tuning curves. (E-F) Similar to C-D but for neuron tuning curves. (G-H) Low-D trajectories of different noise levels correspond to an example image in the primate visual cortex (G) and model (H). The trajectory is averaged across noise patterns. The black arrow indicates the direction along which the noise level changes (denoted as the image-to-noise axis). Triangle: trial start; cross: trial end.}
    \label{fig:manifold_sup}
\end{figure}

As in fig.\ref{fig:manifold_sup}A, the evolution of single-neuron activity has a similar pattern as in Chapter I, despite that the activity of some neurons exhibits multiple bumps (fig.\ref{fig:manifold_sup}A, left panel). Thus, their activities need a longer time to reach steady states. By contrast, in the pre-trained network, the neuron activities will not exhibit multiple bumps (fig.\ref{fig:manifold_sup}A). The more complicated response pattern results from the complex connectivity patterns learned by familiarity training.
The familiarity suppression and sharpening of stimulus tuning and population tuning are also observed in the noise experiment (fig.\ref{fig:manifold_sup}B, C-F) \todo{fig 6 delete A, keep B, C, D + example tuning curves, del E, F, keep G, H}. The evolution of the population activity of excitatory neurons forms a trajectory in the high-dimensional neural space. In both primate neural data and our model, population vectors corresponding to different noise levels start at similar initial points, diverge into separate trajectories, and eventually stabilize to form the attractor that represents distinct input information (fig.\ref{fig:manifold_sup}G-H). Fixed points of different noise levels are arranged along certain directions in the state space (fig.\ref{fig:manifold_sup}G-H), forming an image-to-noise axis along which the signal and noise content change gradually.

\subsection{Additional visualization of the manifold}
\begin{figure}[ht!]
    \centering
    \centerline{\includegraphics[width=\linewidth]{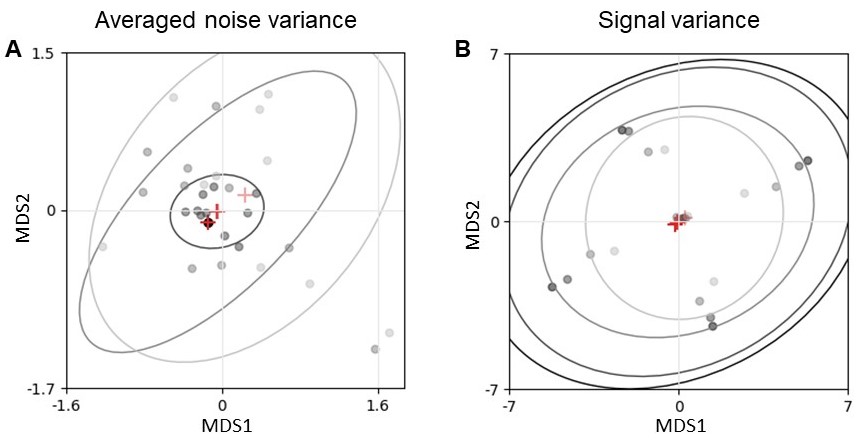}}
    % \captionsetup{width=.85\paperwidth}
    \caption{Manifold visualization. (A) Visualization of the clusters of different noise levels averaged across different images in epoch5 trained network. Dots: noise patterns. Deeper color means a lower noise level. Red crosses: centroids of averaged clusters. Ellipses: covariance of averaged clusters. The color code is identical to the dots. (B) Visualization of the manifold formed by all cluster centroids in epoch5 trained network. Dots: centroids. Deeper color means a lower noise level. Red crosses: average of centroids of the same noise level. Ellipses: covariance of centroids of the same noise level. The color code of ellipses is identical to the dots.}
    \label{fig:manifold_sup2}
\end{figure}

In this section, we visualize the manifold formed by the network's response. For a single image, the manifold formed by all its associated noisy counterparts has approximately a cone shape as expected. Specifically, noise patterns in each noise level form a cluster. The clear image is located at one end of the manifold, and the increasing noise levels would drive the cluster farther away from the clear image. Furthermore, the dispersion of the cluster (noise variance) would be larger for higher noise levels (fig.\ref{fig:manifold_sup2}A). Such geometry arises because the higher noise level reduces the similarity between images corrupted by different noise patterns. Considering all images, we found that as the noise level increased, the mean distances between clusters corresponding to different images (signal variance) decreased. For 100\% noises, all the "images" will converge to the same cloud (data not shown), reflecting the reduced signal contents in the corrupted image. Therefore, if only the cluster centroids are considered, the manifold's shape is also a cone but has the opposite direction compared to the first cone (fig.\ref{fig:manifold_sup2}B).

\section{Aligning slow modes with invariant features in noise experiment}
\subsection{Experiment details}
To illustrate different system behavior when slow/fast modes are aligned with input difference/invariant features, a two-dimensional linear system was constructed, where the dynamic of response $\bm{r}$ was given by: $\partial \bm{r} / \partial t = \bm{M} \bm{r} + I$. The two eigenvectors of $\bm{M}$ were generated by rotating the input difference vector by different angles: $\bm{m_i} = [\cos{\theta_i}, -\sin(\theta_i); \sin(\theta_i), \cos(\theta_i)] (I_1 - I_2)$, with $\theta_1 = 0.01\pi + \pi/2$, $\theta_2 = 0.01\pi$. Eigenvalues $\lambda_i$ of $\bm{M}$ were both real and negative, making the fixed point stable and non-oscillating, and $\tau_i = 1 / \lambda_i$. Then $\bm{M} = U \Lambda U^T$, where $U = [\bm{m_1}, \bm{m_2}]$, and $\Lambda = diag([\lambda_1, \lambda_2])$. 
%In fig.\ref{fig:theory}B, D, the perturbed response was computed using the solution: $\bm{r}(t) = e^{At} \delta r + \bm{r}_{SS}$, where $r_{SS} = -A^{-1} I$, and $\delta r$ was $[-20; 104]$. 
In fig.\ref{fig:theory}D, E, we set $I_1 = [0.1; 1.1]$, $I_2 = [0.5; 1]$, and the response trace was obtained by Euler integration with $dt = 0.2$. Initial state $\bm{r}_0$ were $[16; 75]$ and $[10; 17]$ for D and E, respectively. 

To analyze dynamical modes and their associated decaying time constants in our neural circuit model, we forward-simulated the system until it converges to find the fixed point and used the numerical method to obtain the eigenvalues and eigenvectors of the Jacobian. Then the projection of input derivatives and response derivatives were $L^T r'(\theta)$ and $\Tilde{L}^T \bm{\alpha}'(\theta)$ as derived in Appendix B, where $\Tilde{L} = L \phi'$, and columns of $L$ are top M slow left eigenvectors of the $J(\theta)$ (M ranges from 1 to 200). The decaying time constant of mode $i$ was: $\tau_i = - 1 / Re(\lambda_i)$, where $\lambda_i$ is the $i^{th}$ eigenvalue of $J$. Due to the complexity and unobservability of the feature space of the natural images, the derivative of the input $\bm{\alpha}'(\theta)$ is inaccessible. Therefore, we used the distance between the corresponding fixed points to approximate the derivative. Specifically, when the linearization was performed at the steady-state response of stimulus $\bm{x}^n_{l,p}$, the noise distance of projected inputs is: 
\begin{equation}
    \|\Delta_l \Tilde{L}^T \bm{\alpha}\|_2 = \frac{1}{h}\, \sqrt{(\Tilde{L}^T \bm{\alpha}(\bm{x}^n_{l+h,p}) - \Tilde{L}^T \bm{\alpha}(\bm{x}^n_{l,p}))^2},
\end{equation}
where $n, l, p$ refers to the image index, noise level, and noise pattern, $h$ denotes the increment of noise level to the adjacent noise level, with a unit of 10\%. The image distance of the projected input is 
\begin{equation}
    \|\Delta_n \Tilde{L}^T \bm{\alpha}\|_2 = \sqrt{\frac{1}{N-1}\, \sum_{i\neq n} (\Tilde{L}^T \bm{\alpha}(\bm{x}^i_{l,p}) - \Tilde{L}^T \bm{\alpha}(\bm{x}^n_{l,p}))^2},
\end{equation}
where $N$ is the number of images. The distances of the response projection and the original response of the excitatory population were computed similarly. % Finally, the non-orthogonality of $\bm{\nu}_1$ was calculated as $\| \bm{\nu}_1^T \bm{\nu}_1 - \text{diag}(\bm{\nu}_1^T \bm{\nu}_1) \|_F\, /\, \sqrt{M^2 - M}$.

\subsection{Effects of subspace dimension}
\begin{figure}[h!]
    \centering
    \centerline{\includegraphics[width=\linewidth]{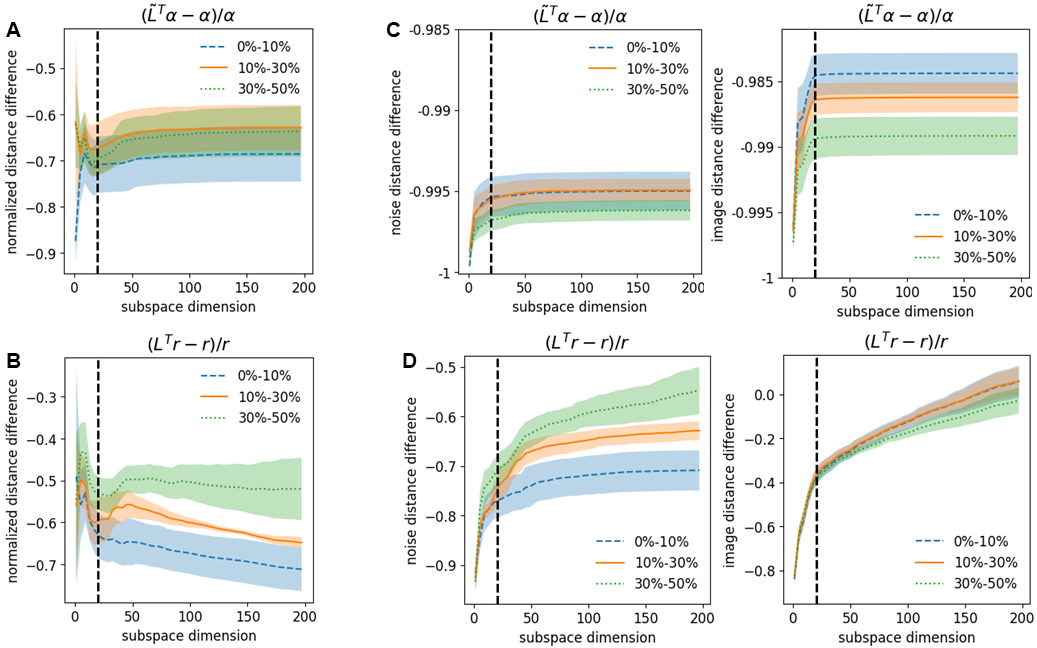}}
    % \captionsetup{width=\textwidth}
    \caption{Relative differences in normalized noise distance, noise distance, and image distance computed in epoch5 trained network. (A) and (C) contrast input projection ($\Tilde{L}^T \alpha$) and input ($\alpha$); (B) and (D) contrast response projection ($L^T r$) and response ($r$). The shaded area represents s.e.m across images. The dashed vertical line indicates the 20-dimensional subspace.}
    \label{fig:dynamic_sup}
\end{figure}

In this section, we extend the analysis of sec.\ref{sec:dynamic} to different subspace dimensions from 1 to 200, containing modes in group 1. We define the alignment with invariant features as the relative difference between normalized noise distance of input $\alpha$ and input projection $\Tilde{L}^T \alpha$, and subspace's response invariance as normalized noise distance of response projection $L^T r$, Fig.\ref{fig:dynamic_sup}A, B show that group 1 dynamical mode consistently detect invariant features, and invariance in the slow modes subspace is consistently better than original response. This may be a result of many modes overlapping with the same feature. Individual distance differences are consistently below zero, with noise distance being smaller than the image distance, and the absolute value of two projected distances reasonably increases as the dimension goes up (fig.\ref{fig:dynamic_sup}C, D). 

\begin{figure}[h!]
    \centering
    \centerline{\includegraphics[width=\linewidth]{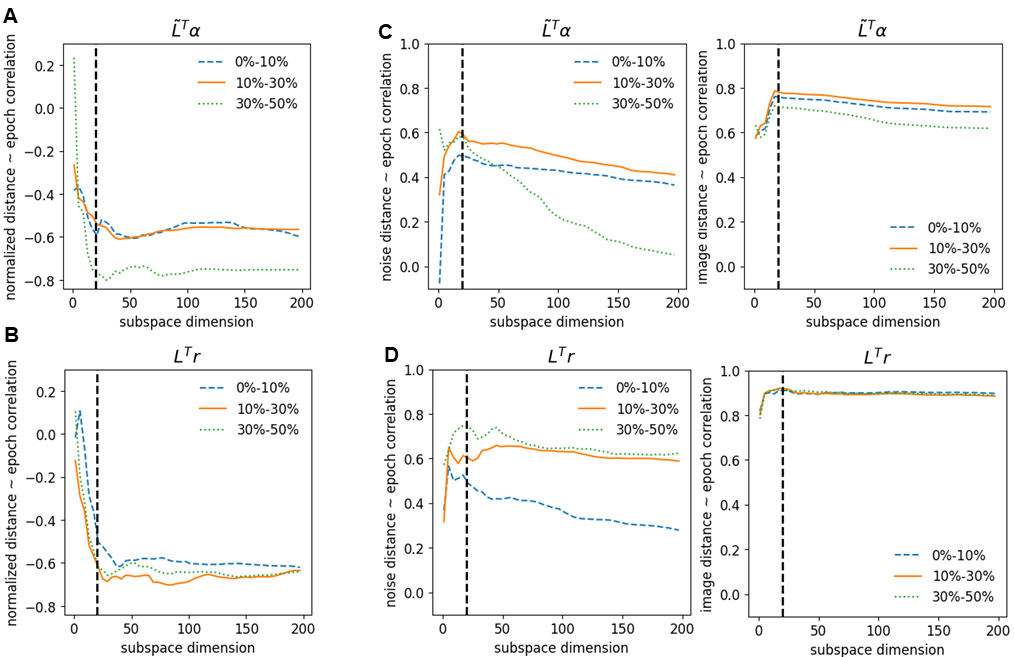}}
    % \captionsetup{width=\textwidth}
    \caption{Training effects of different subspace dimensions. (A-D) Pearson correlation between training epochs and normalized noise distance, noise distance, image distance of input projection ($\Tilde{L}^T \alpha$, top row), and response projection ($L^T r$, bottom row). The dashed vertical line indicates the 20-dimensional subspace.}
    \label{fig:dynamic_sup2}
\end{figure}

To quantify the effect of training, we calculated the Pearson correlation between training epochs and the corresponding quantities for each subspace dimension. Figure.\ref{fig:dynamic_sup2}A, B show that the training effect increases and is saturated with higher subspace dimension. Note that the subspace's response invariance exhibits little training effect when the subspace dimension is low (fig.\ref{fig:dynamic_sup2}B), as the increased alignment is canceled off by the increased time constants (fig.\ref{fig:dynamic}B). For individual projected distances, the training effect is similar across subspace dimensions as described in sec.\ref{sec:dynamic}, and it peaks at around 20 dimensions (fig.\ref{fig:dynamic}C, D). Overall, modes' properties and training effects do not vary much with different subspace dimensions.
% Furthermore, training also increases the non-orthogonality of the 20 most slowly decaying dynamical modes (fig.\ref{fig:dynamic}H, I). It suggests that the invariance in the detected features is non-trivial as the network learned more specific, disentangled features in the images.

% \begin{figure}[h!]
%    \centering
%    \centerline{\includegraphics[width=\textwidth]{images/dynamic_sup3.png}}
%    % \captionsetup{width=\textwidth}
%    \caption{TTTT}
%    \label{fig:dynamic_sup3}
% \end{figure}

\section{Broader impacts}
This is a computational neuroscience study investigating the neural plasticity mechanisms underlying learning in the visual systems. Such fundamental knowledge is important for understanding how the brain works and developing clinical therapy for the visually impaired.  Such knowledge can potentially advance the development of artificial intelligence and deep learning, which would have both positive and negative societal impacts. The positive result would facilitate the development of brain-like visual systems that can learn to quickly adapt to new environments and remember global image context.  While deep learning can advance science and technology, it comes with certain inherent risks to society. We acknowledge the importance of ethical study in all works related to biological and machine learning.  A better understanding of deep learning and the brain, however, is also crucial for combating the misuse of deep learning by bad actors in this technological arms race.

% \newpage

% \input{sections/checklist}

\end{document}